\documentstyle[12pt,fleqn]{article}
\def\singlespace {\smallskipamount=3.75pt plus1pt minus1pt
                  \medskipamount=7.5pt plus2pt minus2pt
                  \bigskipamount=15pt plus4pt minus4pt
                  \normalbaselineskip=15pt plus0pt minus0pt
                  \normallineskip=1pt
                  \normallineskiplimit=0pt
                  \jot=3.75pt
                  {\def\smallskip {\vskip\smallskipamount}}
                  {\def\medskip   {\vskip\medskipamount}}
                  {\def\bigskip   {\vskip\bigskipamount}}
                  {\setbox\strutbox=\hbox{\vrule 
                    height10.5pt depth4.5pt width 0pt}}
                  \parskip 7.5pt
                  \normalbaselines}
\def\middlespace {\smallskipamount=5.625pt plus1.5pt minus1.5pt
                  \medskipamount=11.25pt plus3pt minus3pt
                  \bigskipamount=22.5pt plus6pt minus6pt
                  \normalbaselineskip=22.5pt plus0pt minus0pt
                  \normallineskip=1pt
                  \normallineskiplimit=0pt
                  \jot=5.625pt
                  {\def\smallskip {\vskip\smallskipamount}}
                  {\def\medskip   {\vskip\medskipamount}}
                  {\def\bigskip   {\vskip\bigskipamount}}
                  {\setbox\strutbox=\hbox{\vrule 
                    height15.75pt depth6.75pt width 0pt}}
                  \parskip 11.25pt
                  \normalbaselines}
\def\doublespace {\smallskipamount=7.5pt plus2pt minus2pt
                  \medskipamount=15pt plus4pt minus4pt
                  \bigskipamount=30pt plus8pt minus8pt
                  \normalbaselineskip=30pt plus0pt minus0pt
                  \normallineskip=2pt
                  \normallineskiplimit=0pt
                  \jot=7.5pt
                  {\def\smallskip {\vskip\smallskipamount}}
                  {\def\medskip   {\vskip\medskipamount}}
                  {\def\bigskip   {\vskip\bigskipamount}}
                  {\setbox\strutbox=\hbox{\vrule 
                    height21.0pt depth9.0pt width 0pt}}
                  \parskip 15.0pt
                  \normalbaselines}

\include{dspace12}
\def\be{\begin{equation}}
\def\ee{\end{equation}}
\def\bea{\begin{eqnarray}}
\def\eea{\end{eqnarray}}
\def\nn{\nonumber}
\def\th{\theta}

\def\lt{\left}
\def\rt{\right}
\def\sect #1{\setcounter{equation}{0}}

\begin{document}
\middlespace

\begin{center}
{\LARGE { Nature of  singularity in Einstein-massless scalar theory
}}
\end{center}
\vspace{0.7in}
\vspace{12pt}
\begin{center}
{\large{ 
K. S. Virbhadra, S. Jhingan and P. S. Joshi\\
Theoretical Astrophysics Group\\
Tata Institute of Fundamental Research\\
Homi Bhabha Road, Colaba, Mumbai  400005, India.\\
}}
\end{center}
\vspace{1.3in}
\begin{abstract}
We study the static and  spherically symmetric  exact solution of the
Einstein-massless scalar equations given by Janis, Newman and Winicour.
We find that this solution satisfies the weak energy condition and has  
strong globally naked singularity.
\end{abstract}
\vspace{1.0in}
\begin{center}
{\em To appear in  Int.  J. Mod. Phys. D}
\end{center}
\newpage

Although the general theory of relativity is one of the most
beautiful physical theory and  is supported by experimental evidences
with great success, there are some questions which are
not yet answered. For instance, Penrose\cite{Pen69}, in a classic paper, asked if
there is a cosmic censor who forbids the occurrence of naked singularities,
clothing each one in an absolute event horizon. 
Unfortunately,
there exists no agreement on a precise statement of  cosmic censorship
hypothesis. However, it is usually stated that while the energy conditions
are satisfied, a globally naked  stable singularity cannot be
produced by a regular initial data.
Whether or not  naked singularities exit in nature is a
very important issue, because many important results in general
relativity are based on the cosmic censorship hypothesis. Moreover,
if the naked singularities exist in nature one might have a
chance to study the effects of highly curved regions of spacetime.
Despite many painstaking efforts, no proof for any version of cosmic
censorship hypothesis is known. One of the main obstacles towards
achieving this goal is the lack of a precise mathematical formulation
describing  ``a physically realistic system''. When a proof
appears difficult it is worth obtaining counterexamples. In this
context,  several authors have studied  gravitational
collapse and the cosmic censorship problem in great detail (see
\cite{Jos93} and references therein).

 It is known that the scalar fields have been suspected for
causing the long-range gravitational fields. Inspired by this, many  theories
have been proposed since before general theory of relativity\cite{Wie72}.
After the Brans-Dicke theory\cite{BD61} was given, the scalar fields minimally
as well as conformally  coupled to gravitation have been a
subject of immense interest to many  researchers (see
\cite{Scalar}-\cite{CScalar} and references therein). Several authors have put considerable
efforts to obtain solutions to the minimally coupled Einstein-massless scalar
(EMS) equations\cite{Scalar}-\cite{JNW68}. In the present paper we consider the  well-known
static and spherically symmetric  exact solution
of EMS, given by Janis, Newman and Winicour(JNW)\cite{JNW68}
and study the nature of singularity in this solution.
We use the geometrized units (the gravitational constant $G=1$, the
 speed of light in vacuum $c=1$) and follow the convention that the Latin indices
run from $0$ to $3$ ($x^0$ is the time coordinate).

The Einstein-massless scalar field equations are 
\be
R_{ij}\ -\ \frac{1}{2}\ R \ g_{ij}\ =\ 8 \pi \ S_{ij}\ ,
\ee
where $S_{ij}$, the energy-momentum tensor of the massless scalar field, is
given by
\be
S_{ij}\ =\ \Phi_{,i}\ \Phi_{,j}\ -\ \frac{1}{2}\ g_{ij}\ g^{ab}\ \Phi_{,a}\
          \Phi_{,b}\ ,
\ee
and 
\be
\Phi_{,i}^{\ ;i}\ =\ 0 \ , 
\ee
where, $\Phi$ stands for the massless scalar field. $R_{ij}$ is
the Ricci tensor and $R$ is the Ricci scalar.
Equation $(1)$ with Eq. $(2)$ can be written as
\be
R_{ij}\ =\ 8 \pi \ \Phi_{,i}\ \Phi_{,j} .
\ee
It is usually assumed that for any physically reasonable
classical matter
\be
T_{ij} V^i V^j \geq 0 ,
\ee
where $T_{ij}$ is the energy-momentum tensor and $V^i$ stands
for the four-velocity of any timelike observer. This is known as
the {\it weak energy condition}\cite{Wal84}. For any massless scalar
field
\be
S_{ij} V^i V^j = \lt( \Phi_{,i} V^i \rt)^2 - \frac{1}{2} (g_{ij}
                 V^i V^j) g^{ab} \Phi_{,a} \Phi_{,b} .
\ee
A sufficient condition for a strong
curvature singularity, as defined by Tipler et al\cite{TCE84}, is that 
 at least along one null geodesic (with an affine parameter
$k$) that ends at singularity  $k=0$, the following is satisfied.
\be
\lim_{k \rightarrow 0} k^2 R_{ij} v^i v^j \neq 0 ,
\ee
where $v^i \equiv \frac{dx^i}{dk}$ is the tangent vector to the
null geodesics\cite{Strength}. For the massless scalar field 
\be
k^2 R_{ij} v^i v^j = 8 \pi k^2 (\Phi_{,i} v^i)^2 .
\ee

JNW\cite{JNW68} obtained  static and spherically symmetric  exact solution of EMS 
equations, which is given by  the line element
\be
ds^2\ =\ B dt^2 - A dr^2 - D r^2 (d\th^2 + \sin^2\th d\phi^2) ,
\ee
with
\bea
B &=& A^{-1} = 
{\lt(
\frac{1-\frac{a_-}{r}}{1+\frac{a_+}{r}}
\rt)}^{1/{\mu}} , \nn\\
D &=& \lt(1-\frac{a_-}{r}\rt)^{1-1/{\mu}}
        \lt(1+\frac{a_+}{r}\rt)^{1+1/{\mu}} ,
\eea
and the scalar field 
\be
\Phi\ =\ \frac{\sigma}{\mu}
\ln \lt(
\frac{1-\frac{a_-}{r}}{1+\frac{a_+}{r}}
\rt) ,
\ee
where
\be
a_{\pm} = r_0 (\mu \pm 1)/2
\ee
and
\be
\mu = \sqrt{1+16 \pi {\sigma}^2} .
\ee

Here $r_0$ and $\sigma$ are two constants in the solution.
$\sigma=0$ gives the Schwarzschild solution and $r_0=0$ gives
the flat spacetime.   The only nonvanishing component of the Ricci tensor is
\be
R_{rr} = \frac{ {r_0}^2 (\mu^2-1)}
              {2 (r+a_+)^2 (r-a_-)^2}.
\ee
The nonvanishing components of the Einstein tensor  are 
\be
G^0_0 = - G^1_ 1 =  G^2_2 =  G^3_3 = 
\frac{ {r_0}^2 (\mu^2-1)   }
     {  4 (r+a_+)^{2+\frac{1}{\mu}}  (r-a_-)^{2-\frac{1}{\mu}} } .
\ee
As the real scalar field $\Phi$ in JNW solution depends only on
the radial coordinate, Eq. (6) becomes
\be
S_{ij} V^i V^j = (\Phi_{,i} V^i)^2 - \frac{1}{2} (g_{ij} V^i
V^j) g^{rr} (\Phi_{,r})^2.
\ee
Thus it is clear that the  weak energy condition is satisfied for the JNW
 spacetime.

It has been a difficult task to determine a curvature singularity in a given
spacetime(\cite{Wal84}-\cite{Ger68}). The curvature singularity is usually found
by showing the divergence of the Kretschmann scalar at a finite
affine parameter along some geodesic. The Ricci as well as the
Weyl scalars are known to be finite for several types of
solutions having curvature singularities. For instance, the
Ricci scalar vanishes for electrovac  solutions
and the Weyl scalar vanishes for any conformally flat spacetime.
In any case, the divergence of any of these scalars at a finite
affine parameter demonstrates the presence of a curvature singularity.

The Ricci, the Kretschmann, and the Weyl scalars for the JNW
metric are, respectively, given by
\be
 R_{ij} g^{ij} = - G_{ij} g^{ij} = 2 G^1_1 ,
\ee
\be
R_{ijkl} R^{ijkl} = 
\frac{ {r_0}^2 (3 \mu^4 {r_0}^2 - 16 \mu^2 r r_0 - 6 \mu^2 {r_0}^2
       + 48 r^2 + 16 r r_0 + 3 {r_0}^2)}
     {  4 (r+a_+)^{4+\frac{2}{\mu}}  (r-a_-)^{4-\frac{2}{\mu}}      }
\ee
and
\be
C_{ijkl} C^{ijkl} = 
\frac{ {r_0}^2 (\mu^2 r_0 - 6 r - r_0)^2}
      {  3 (r+a_+)^{4+\frac{2}{\mu}}  (r-a_-)^{4-\frac{2}{\mu}}}.
\ee
$R_{ijkl}$ and $C_{ijkl}$ are, respectively, the Riemann
curvature and the Weyl tensors.
It is obvious that these scalars diverge at $r=a_-$ showing a curvature
singularity there. Now we investigate whether or not this singularity
is naked. A singularity is called globally naked if there is a future
directed causal curve with one end ``on the singularity'' and the other end 
on the future null infinity. The null geodesics equations are
\be
\frac{dv^i}{dk} + \Gamma^i_{jk} v^j v^k = 0,
\ee
where
\be
g_{ij} v^i v^j = 0.
\ee
The outgoing radial null geodesics in JNW spacetime are given by
\bea
v^t &=& \frac{E}{B}, \nn\\
v^r &=&  E, \nn\\
v^{\th} &=& v^{\phi} = 0,
\eea
where $E (E>0)$ is an integration constant.
Thus one has
\be
r =  k  E  + a_-  
\ee
and
\be
dt =    {\lt(\frac{r+a_+}{r-a_-}\rt)}^{\beta} dr,
\ee
where $\beta = 1/\mu$. 
For finite $R$,
\bea
\lim_{\epsilon \rightarrow 0} \ 
 \int^R _{a_{-} + \epsilon}\ {\lt(\frac{r+a_{+}}{r-a_{-}}\rt)}^{\beta} dr
&<& \lim_{\epsilon \rightarrow 0} \ 
(R+a_{+})^{\beta} \int^R _{a_{-} + \epsilon} \  \frac{dr}
{\lt(r- a_{-} \rt)^{\beta} }\nn\\
&=& (R+a_{+})^{\beta} \  \frac{\lt(R-a_{-} \rt)^{1 - \beta}}{\lt(1 - \beta\rt)},
\eea
which is  finite.
Thus, the singularity $r=a_-$ is globally
naked for all values of the constant parameters $r_0$ and $\mu$
of the JNW solution. Further, as
\be
\lim_{k \rightarrow 0} k^2 R_{ij} v^i v^j = \frac{1-\beta^2}{2}
\ee
is clearly nonvanishing, the globally naked
singularity in JNW solution is a strong curvature singularity.

The cosmic censorship hypothesis has been debated (see Joshi \cite{Jos93} and
 references therein). Penrose\cite{Pen72} suggested  that the possibility
 that naked singularities may sometimes arise must be considered seriously.
In the present paper we have shown that the  singularity
in the  JNW solution is globally naked strong curvature
singularity. 
However, it remains to be investigated whether or not this naked
singularity occurs in the collapse from a  reasonable nonsingular
initial data. Without that, it  may not be taken  as  a serious counterexample to 
the cosmic  censorship.
It is also of interest to study whether or not the  naked singularity
discussed here is stable.
\begin{flushleft}
{\bf Acknowledgements}
\end{flushleft}
Thanks are due to T P Singh and I H  Dwivedi for helpful discussions.
\newpage


\begin{thebibliography}{99}
\setlength{\parskip}{0.32ex}

\bibitem{Pen69}
       R. Penrose, {\em Riv. del Nuovo Cim.}  {\bf 1},  252 (1969).
\bibitem{Jos93}
       P. S. Joshi,     Global aspects in  gravitation and cosmology,
          (Clarendon Press, Oxford, 1993);
       P. S. Joshi and T. P. Singh, {\em  Phys. Rev.}  {\bf D51},  6778 (1995);
       K. S. Virbhadra,  gr-qc/9606004.
\bibitem{Wie72}
       S. Weinberg,     Gravitation and cosmology: principles
          and applications of the general  theory of relativity
          (John Wiely \& Sons, NY, 1972) p.157.
\bibitem{BD61}
       C. H. Brans and R. H. Dicke,  {\em  Phys. Rev.}  {\bf 124},  925 (1961).
\bibitem{Scalar}
       R. Penny,  {\em  Phys. Rev.}  {\bf 174},   1578 (1968);
       A. I. Janis, D. C. Robinson  and J. Winicour, {\em Phys. Rev.} {\bf 186},
           1729 (1969);
       H. A. Buchdahl, {\em  Gen. Rel. Grav.}  {\bf 9},  59 (1978); 
       B. C. Xanthopoulos and T. Zannias, {\em  Phys. Rev.}  {\bf D40}, 
           2564 (1989);
       V. Husain, E. A.  Martinez and D. N\'{u}\~{n}ez, {\em  Phys. Rev. } {\bf
          D50},   3783 (1994); 
       K. S. Virbhadra, gr-qc/9408035, {\em Pramana- J.Phys.}  {\bf 44},  (1995) 317.
\bibitem{JNW68}
     A. I. Janis, E. T. Newman and J. Winicour,  {\em Phys. Rev. Lett.}  
          {\bf 20},   878 (1968).
\bibitem{CScalar}
       C. G. Callan Jr., S. Coleman and R. Jackiw, {\em Ann. Phys.(NY)} {\bf 59},
           42 (1970); 
       L. Parker,   {\em Phys. Rev.}  {\bf D7},   976 (1973);
       J. D. Bekenstein,  {\em  Ann. Phys.}  {\bf 82},  535 (1974);  
                          {\em  Ann.  Phys.}  {\bf 91},  75 (1975);
       B. C. Xanthopoulos, {\em  J. Math. Phys.}  {\bf 32},  1875 (1991);
       K. S. Virbhadra and J. C. Parikh, {\em Phys. Lett.} {\bf B331},  302 (1994);
                               Erratum : {\em Phys. Lett.} {\bf B340},  265 (1994).
\bibitem{Wal84}
       R. M. Wald, General Relativity (The University of Chicago Press,
      Chicago, 1984) p.219, p.148.
\bibitem{TCE84}
       F. J. Tipler, C. J. S. Clarke and G. F. R. Ellis, in {\em
              General Relativity  and Gravitation},  edited by A. Held 
              (Plenum, NY, 1980) p.97.
\bibitem{Strength}
        B. Waugh and K. Lake, {\em Phys. Rev.} {\bf D38}, 1315 (1988);
        A. Ori and T. Piran,  {\em  Phys. Rev.} {\bf D42},  1068 (1990); 
      C. J. S. Clarke and A. Kr\'{o}lak, {\em J.  Geom. Phys.} {\bf 2}, 127 (1986).
\bibitem{Ger68}
       R. P. Geroch,  {\em  Ann. Phys.}  {\bf 48},   526 (1968).
\bibitem{Pen72}
        R. Penrose,  {\em  Nature} {\bf 236},    377 (1972).


\end{thebibliography}
\end{document}